\documentclass[prd,twocolumn,showpacs,preprintnumbers,superscriptaddress,amsmath,amssymb,nofootinbib]{revtex4}

\usepackage[dvips,xdvi]{graphicx,psfrag}
\usepackage{dcolumn}
\usepackage{bm}
\usepackage{amsmath, amssymb}

\newcommand{\Psfig}[2]{\includegraphics[width=#1]{#2}}
\newcommand{\Expect}[1]{\langle #1 \rangle}
\newcommand{\SUN}[1]{\text{SU} ( #1 )}
\newcommand{\UN}[1]{\text{U} ( #1 )}

\def\bc{\begin{center}}
\def\ec{\end{center}}

\def\prt{\partial}
\def\trace{\text{tr}}

\usepackage[normalem]{ulem}  
\usepackage[dvips]{color} 

\renewcommand\sout{\bgroup \color{red} \ULdepth=-.5ex \ULset}
\begin{document}

\preprint{}

\title{New Approach to Continuum Path Integrals for Particles and Fields}

\author{Takayasu Sekihara}
\altaffiliation[Present address: ]{Institute of Particle and Nuclear
  Studies, High Energy Accelerator Research Organization (KEK), 1-1,
  Oho, Ibaraki 305-0801,
  Japan.}
\affiliation{Department of Physics, Tokyo Institute of Technology, 
  Tokyo 152-8551, Japan}

\date{\today}

\begin{abstract}
  An approach to approximate evaluation of the continuum Feynman path
  integrals is developed for the study of quantum fluctuations of
  particles and fields in Euclidean time-space.  The paths are
  described by sum of Gauss functions and are weighted with $\exp
  (-S)$ by the Metropolis method.  The weighted smooth paths reproduce
  properties of the ground state of the harmonic oscillator in one
  dimension with more than about $90 \%$ accuracy, and the accuracy
  gets higher by using smaller width of the Gauss functions.  Our
  approach is applied to quantum field theories and quantum
  fluctuations of U(1) and SU(2) gauge fields in four dimensions
  respectively provide the Coulomb force and confining linear
  potential at qualitative levels via the Wilson loops.  Distributions
  of large values of gauge fields are found to be suppressed at least
  exponentially.
\end{abstract}


\maketitle


It is quantum physics which dominates microscopic phenomena less than
the atomic scale~\cite{Sakurai:2010mqm}.  One elegant way to describe
the quantum phenomena is the path integral method developed by
Feynman~\cite{Feynman:1948ur}, in which all possible paths are taken
into account with the probability amplitude $\exp (i S/\hbar)$ with
$S$ the action of the system.  The path integral method gives a clear
interpretation of behavior of particles in quantum mechanics as
fluctuations from the classical paths, although exact evaluations of
the path integrals are possible only in few cases such as the harmonic
oscillator~\cite{Feynman:2010QM}.  The path integral method also
promotes modern developments of theoretical elementary particle
physics and supplies a nonperturbative technique for quantum field
theories~\cite{Weinberg:2005QFT}.

The evaluation of the path integrals can be simplified by discretizing
time-space, in which derivations and integrations are replaced with
finite differences and summations, respectively, and measure of the
path integrals becomes a countable product.  An important example of
the discretized path integrals is the lattice QCD (quantum
chromodynamics)~\cite{Wilson:1974sk}, by which nonperturbative aspects
of QCD have been revealed~\cite{Rothe:2005LGT}.  However, the
time-space discretization explicitly breaks continuous symmetries of
time-space such as the translational symmetry down to discrete
symmetries, and sometimes leads to qualitative discrepancies such as
magnetic monopoles in the lattice QED (quantum
electrodynamics)~\cite{Polyakov:1975rs}.  Hence it is desired to
perform the path integrals in continuous time-space from viewpoint of
complementarity for the discretized approach.

In this paper we develop an approach to evaluating the continuum path
integrals in Euclidean time ($t \to - i \tau$) for particles and
fields.  The continuous paths are described by sum of smooth functions
with weight $\exp (-S)$ by the Metropolis
method~\cite{Metropolis:1953am}.  Here we take the natural units
$\hbar = c = k_{\rm B}= 1$.


Firstly, for a nonrelativistic particle with one degree of freedom $q$
in a periodic boundary condition with period ${\cal T}$, $q(\tau +
{\cal T}) = q(\tau )$, the path integral method evaluates the quantum
transition amplitude in Euclidean time as~\cite{Feynman:1948ur},
\begin{equation}
{\cal Z} = \int _{\rm period} {\cal D} q \exp ( - S [ q ] ) ,  
\quad 
{\cal D} q \equiv \prod _{\tau} d q (\tau) . 
\label{eq:PI-Euclid}
\end{equation}
Here the measure ${\cal D} q$ is formally defined as an uncountable
product, and the expression~\eqref{eq:PI-Euclid} means that the
quantum transition amplitude corresponds to the summation of all
possible paths for the particle with the probability amplitude $\exp
(-S)$.  Since the quantum fluctuations of the particle are weighted
with the factor $\exp (-S)$, an expectation value of an operator
${\cal O}[q]$ in quantum mechanics can be evaluated by using $N$ paths
$q_{n}$ ($n=1,\, 2, \, \cdots , \, N$) weighted with $\exp (-S)$ as,
\begin{equation}
\Expect{{\cal O} [q]} 
= \frac{1}{\cal Z} \int _{\rm period} 
{\cal D} q {\cal O} [q] \exp ( - S )
\approx 
\frac{1}{N} \sum _{n=1}^{N} {\cal O} [ q_{n} ] , 
\end{equation}
where the last approximation becomes good for large $N$.  

Before explaining our approach to the continuum path integrals, we
briefly review the discretized approach to the path integrals for a
nonrelativistic particle in the periodic boundary condition.  In this
case the particle position is represented as $\tilde{q}_{i}$ at time
$\tilde{\tau} _{i} = i a$ with $i=1$, $\cdots$, $N_{\rm lat}$ and the
lattice spacing $a \equiv {\cal T} / N_{\rm lat}$, and the path of the
particle is obtained by connecting $\tilde{q}_{i}$ and
$\tilde{q}_{i+1}$ from $i=1$ to $N_{\rm lat}$ with straight lines.
Then measure of the path integral is defined as a countable product,
\begin{equation} 
  {\cal D} q \equiv
  \prod _{i=1}^{N_{\rm lat}} d \tilde{q}_{i} ,
  \label{eq:measure}
\end{equation}
and the action is replaced with the corresponding discretized one,
$\tilde{S} [\tilde{q}]$.  The discretized path integrals are evaluated
in the following way (see~\cite{Creutz:1980gp}).  Namely, change of
the particle position at each time $\tilde{\tau}_{i}$, $\delta
\tilde{q}_{i}$, is generated as a random number within $[ - \Delta ,
\, \Delta]$ with a fixed value $\Delta$, in which $\delta
\tilde{q}_{i}$ has a property of the micro-reversibility.  This change
$\delta \tilde{q}_{i}$ is judged by the Metropolis
test~\cite{Metropolis:1953am}, in which $\tilde{q}_{i} + \delta
\tilde{q}_{i}$ is redefined as $\tilde{q}_{i}$ in acceptance
probability $\text{min} [ 1, \, \exp
(\tilde{S}[\tilde{q}]-\tilde{S}[\tilde{q}+\delta \tilde{q}])]$ and
otherwise $\delta \tilde{q}_{i}$ is rejected.  We denote this step as
$W_{i}$.  Then the whole positions are updated by the ``sweep'', {\it
  i.e.}, performing $W_{i}$ from $i=1$ to $N_{\rm lat}$.  After
several sweeps quantum paths in equilibrium weighted with $\exp
(-\tilde{S})$ are obtained.

Now let us make our procedure for the simulation of the continuum path
integrals from analogy to the discretized approach.  Our idea here is
to connect the neighboring points for the particle positions by smooth
lines rather than straight lines.  In order to achieve this, we smear
the particle position in the discretized approach $\tilde{q}_{i}$ with
a Gauss function of width $\xi _{i}$ at time $\tau _{i}$,
\begin{equation}
  \tilde{q}_{i} ~ (\text{at } \tau = \tilde{\tau} _{i}) ~~ \to ~~
  q_{i} \exp \left [ - \frac{(\tau - \tau _{i})^{2}}
    {\xi _{i}^{2}} \right ] ,
\end{equation}
for $i=1$ to $N_{\rm lat}$, where $\tau - \tau _{i}$ means to take
time distance between $\tau$ and $\tau _{i}$ in the periodic boundary
condition.  Here ($q_{i}$, $\tau _{i}$, $\xi_{i}$) in the continuous
approach corresponds to ($\tilde{q}_{i}$, $\tilde{\tau} _{i}$, $a$) in
the discretized approach.  With this smearing one can naturally
connect the path with smooth lines rather than straight lines as,
\begin{equation}
  q (\tau ) 
  = \sum _{i=1}^{N_{\rm sum}} q_{i} 
  \exp \left [ - \frac{(\tau - \tau _{i})^{2}}
    {\xi _{i}^{2}} \right ] , 
  \label{eq:weighted-path}
\end{equation}
where $N_{\rm sum} (=N_{\rm lat})$ is number of the summed terms.  In
this construction of the smooth path, as $q_{i}$ for each $i$ takes
value in range $[-\Lambda , \, \Lambda]$ with a some cut-off $\Lambda$
according to Eq.~\eqref{eq:measure}, $q(\tau)$ at each time takes
value in similar range $[-\Lambda ^{\prime} , \, \Lambda ^{\prime}]$
with a cut-off $\Lambda ^{\prime}$ similar to $\Lambda$, and the
minimal scale of the fluctuation for $q(\tau)$ corresponds to $\xi
_{i}$.  The positions of the Gauss functions $\tau _{i}$ may take
random values rather than values in same interval, $\tau _{i}= i {\cal
  T} / N_{\rm sum}$, as long as every time is equally treated without
making any special time.  The width of the Gauss function, or the
scaling constant, $\xi _{i}$, is fixed so that $\xi _{i}$ does not
depend on $\tau _{i}$, which prevents any special time.  In this study
we take two strategies for $\xi _{i}$; one is to generate $\xi _{i}$
randomly within $[\lambda _{\xi}, \, \Lambda _{\xi}]$ in uniform
probability with ultraviolet and infrared cut-offs $\lambda _{\xi}$
and $\Lambda _{\xi}$, respectively (random scale), and the other is to
use a properly fixed value (fixed scale).  Then the weight $\exp (-S)$
is given by the Metropolis method for $q_{i}$ as in the discretized
approach.  Namely, an additional fluctuation $\delta q_{i}$ with
randomly chosen $i$ is randomly generated within $[-\Lambda _{q},
\Lambda _{q}]$ in uniform probability, where $\Lambda _{q}$ is a
cut-off for the fluctuation amplitude. Then the charge $\delta q_{i}$
with respect to $q_{i}$ is judged by the Metropolis test, in which
$q_{i} + \delta q_{i}$ is redefined as $q_{j}$ in acceptance
probability $\text{min} [ 1, \, \exp (S[q]-S[q+\delta q])]$ and
otherwise $\delta q_{i}$ is rejected.  We emphasize that the
additional fluctuation $\delta q_{i}$ is micro-reversible without
making any special directions in coordinate space.

Here we note that $q (\tau)$ in Eq.~\eqref{eq:weighted-path} cannot
describe all possible paths, since the Gauss functions in
Eq.~\eqref{eq:weighted-path} cannot be a complete set with respect to
the smooth functions in the periodic boundary condition.  Therefore,
at this point our construction of paths is an approximation with
respect to the complete paths required by the measure ${\cal D}q$.
Nevertheless, we expect that most of the possible paths can be taken
into account when value of $N_{\rm sum}$ is sufficiently large with
large $\Lambda _{q}$, small $\lambda _{\xi}$, large $\Lambda _{\xi}$,
and dense $\tau _{i}$, in which case distribution of the Gauss
functions with various width is sufficiently dense.  We also note that
time uniformity may be broken when the time components $\tau _{i}$
take values in same interval, $\tau _{i}= i {\cal T} / N_{\rm sum}$,
but we expect that the uniformity will restore if one considers
sufficiently dense distribution of the Gauss functions.

Our procedure can be summarized as follows: 
\begin{enumerate}

\item Make an initial condition for the path~\eqref{eq:weighted-path}
  by determining constants ($q_{i}$, $\tau _{i}$, $\xi_{i}$) from
  $i=1$ to $N_{\rm sum}$ in the following manner.  Namely, $\tau _{i}$
  is generated within range $[0, \, {\cal T}]$ so as not to make any
  special time, and $\xi _{i}$ is randomly generated within $[\lambda
  _{\xi}, \, \Lambda _{\xi}]$ in uniform probability or is fixed as a
  proper value.  The coefficient $q_{i}$ is randomly generated within
  $[- \Lambda _{q}, \, \Lambda _{q}]$ (hot start) or is taken to be
  zero for all $i$ (cold start).

\item Randomly choose $i$ and generate $\delta q_{i}$ within $[-
  \Lambda _{q}, \, \Lambda _{q}]$ in uniform probability so as to
  construct an additional fluctuation,
  \begin{equation}
    \delta q ( \tau ) = \delta q_{i} \exp \left [ - 
      \frac{(\tau - \tau _{i})^{2}}{\xi _{i} ^{2}} \right ] . 
  \end{equation}

\item According to the Metropolis method~\cite{Metropolis:1953am},
  accept the additional fluctuation $\delta q$ in probability $
  \text{min} [ 1, \, \exp (S[q]-S[q+\delta q])]$.  If and only if the
  additional fluctuation $\delta q$ is accepted, we redefine the path
  $q+\delta q$ as $q$ (or equivalently redefine the coefficient $q_{i}
  + \delta q_{i}$ as $q_{i}$).

\item Iterate steps 2. and 3. until the action as well as other
  expectation values converge.

\end{enumerate}
In this procedure, we eventually obtain a smooth path for the
particle~\eqref{eq:weighted-path}, which is weighted with $\exp (-S)$
due to the step 3.  Generalization to the systems with $f$ degrees of
freedom ($f\ge 2$) is obvious.  Here we emphasize that the description
of smooth fluctuations only by the finite terms of Gauss functions is
an approximation.


Now let us examine our approach by investigating a harmonic oscillator
in one dimension, which action is written as,
\begin{equation}
S_{\rm HO} = \int _{0}^{\cal T} d \tau L_{\rm HO} (q , \, \dot{q} ) , 
\quad 
L_{\rm HO} = \frac{1}{2} m \dot{q}^{2}  
+ \frac{1}{2} m \omega ^{2} q^{2} , 
\label{eq:HOaction}
\end{equation}
with $\dot{q}\equiv dq/d\tau$.  Here we fix its mass and angular
frequency as $m = \omega = 1$, and we take the time range ${\cal T}
=20$.  We consider three simulation conditions; $N_{\rm sum}=50$,
$\Lambda _{q}=3$, and the fixed scale with $\xi = 1$ (A), $N_{\rm
  sum}=100$, $\Lambda _{q}=3$, and the random scale with $\lambda
_{\xi}=0.2$ and $\Lambda _{\xi}=1$ (B), $N_{\rm sum}=200$, $\Lambda
_{q}=3$, and the fixed scale with $\xi = 0.2$ (C).  For the initial
condition we fix $\tau _{i}$ as $\tau _{i}= i {\cal T} / N_{\rm sum}$
and randomly generate $q_{i}$ within $[-\Lambda _{q}, \Lambda _{q}]$
as a hot start.  We prepare $N=400$ paths for the three conditions,
respectively.  Since temperature of the system $1/{\cal T}$ is much
smaller than the excitation energy, the quantum fluctuations in this
study will reflect the ground state of the harmonic oscillator.

\begin{figure}[t]
  \centering
  \begin{tabular}{c}
    \Psfig{8.6cm}{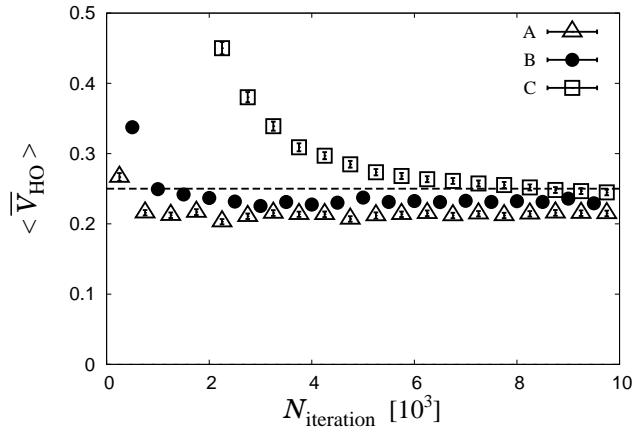} 
  \end{tabular}
  \caption{Expectation values of averaged potential term for the
    harmonic oscillator in the conditions A, B, and C (see text).
    Dashed line denotes the potential expectation value of the ground
    state ($\omega /4$). }
  \label{fig:HO_pot}
\end{figure}

As a result of the numerical simulation, quantum paths approach to
equilibrium at around the iteration number, {\it i.e.}, number of
steps 2--3., $N_{\rm iteration}\approx 10^{3}$ for the condition A,
$N_{\rm iteration}\approx 3 \times 10^{3}$ for B, and $N_{\rm
  iteration}\approx 9 \times 10^{3}$ for C.  In Fig.~\ref{fig:HO_pot}
we show cooling behavior of the potential expectation value $V_{\rm
  HO} \equiv m \omega ^{2} q^{2}/2$ in averaged form
[$\overline{A}\equiv \int _{0}^{\cal T} d \tau A (\tau ) / {\cal T}$]
for the harmonic oscillator by the Metropolis method.  As one can see,
at the saturation point the potential expectation value reproduces the
ground-state value ($=\omega /4$) with more than about $90 \%$
accuracy in all conditions A, B, and C.  Indeed, after iteration
$N_{\rm iteration}=10^{4}$ the expectation value becomes
$\Expect{\overline{V}_{\rm HO}}= 0.213 \pm
0.04$ 
for the condition A, $0.237 \pm 0.004$ 
for B, and $0.241 \pm 0.004$ 
for C.  The potential expectation value approaches to the ground-state
value as the scale constant $\xi$ gets small and hence structures of
the quantum fluctuations in fine scale can be described, which is
similar to the case of the discretized approach.  We expect that the
accuracy of the potential expectation value will get higher if one
uses even smaller $\xi$, although in which case large $N_{\rm sum}$,
large $N_{\rm iteration}$, and large simulation time are required.  We
note that the expectation values are saturated with respect to $N_{\rm
  sum}$, namely we obtain same expectation values within range of
statistical errors with larger $N_{\rm sum}$ in each conditions.

\begin{figure}[t]
  \centering
  \begin{tabular}{c}
    \Psfig{8.6cm}{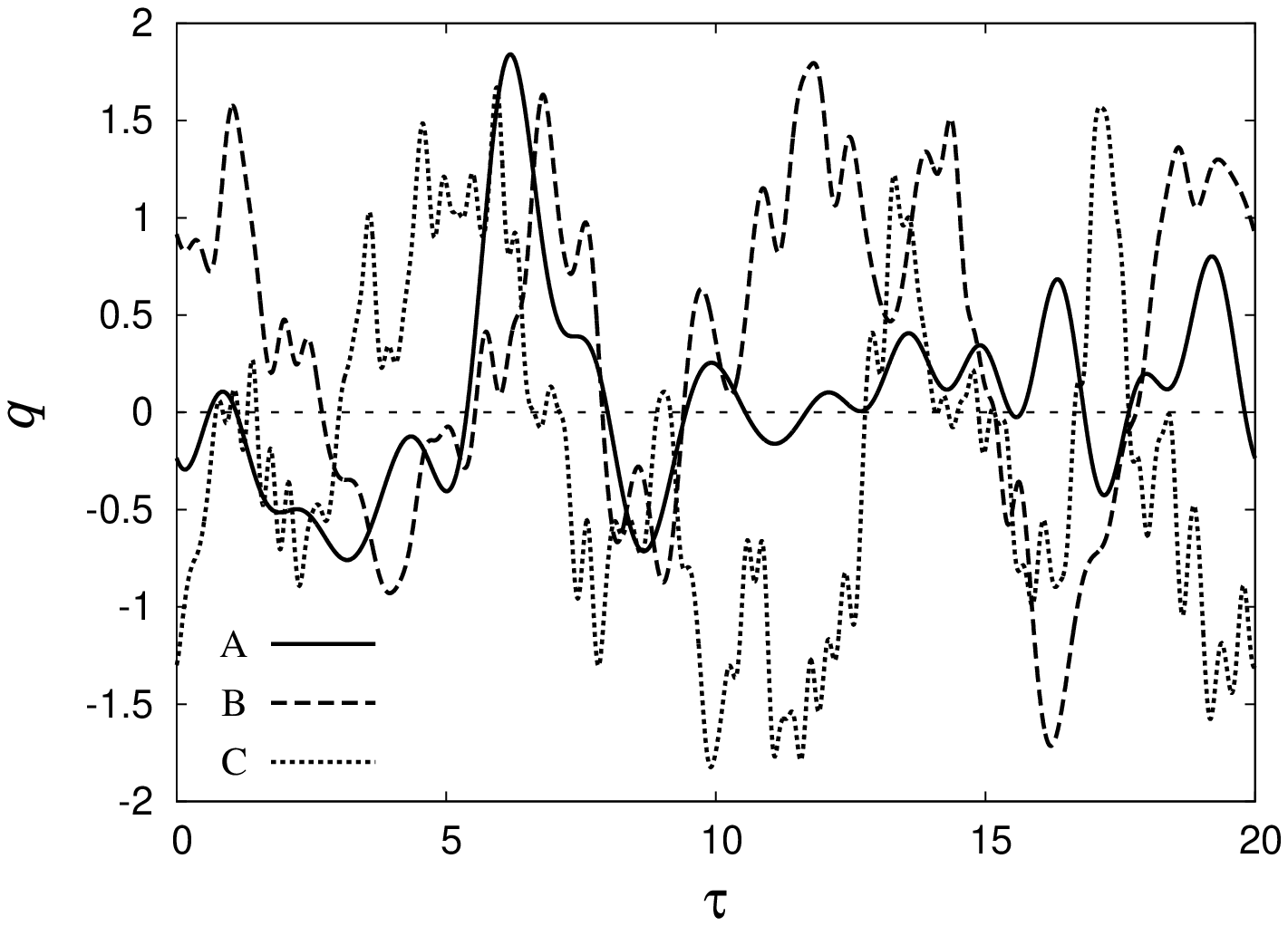} \\
    \Psfig{8.6cm}{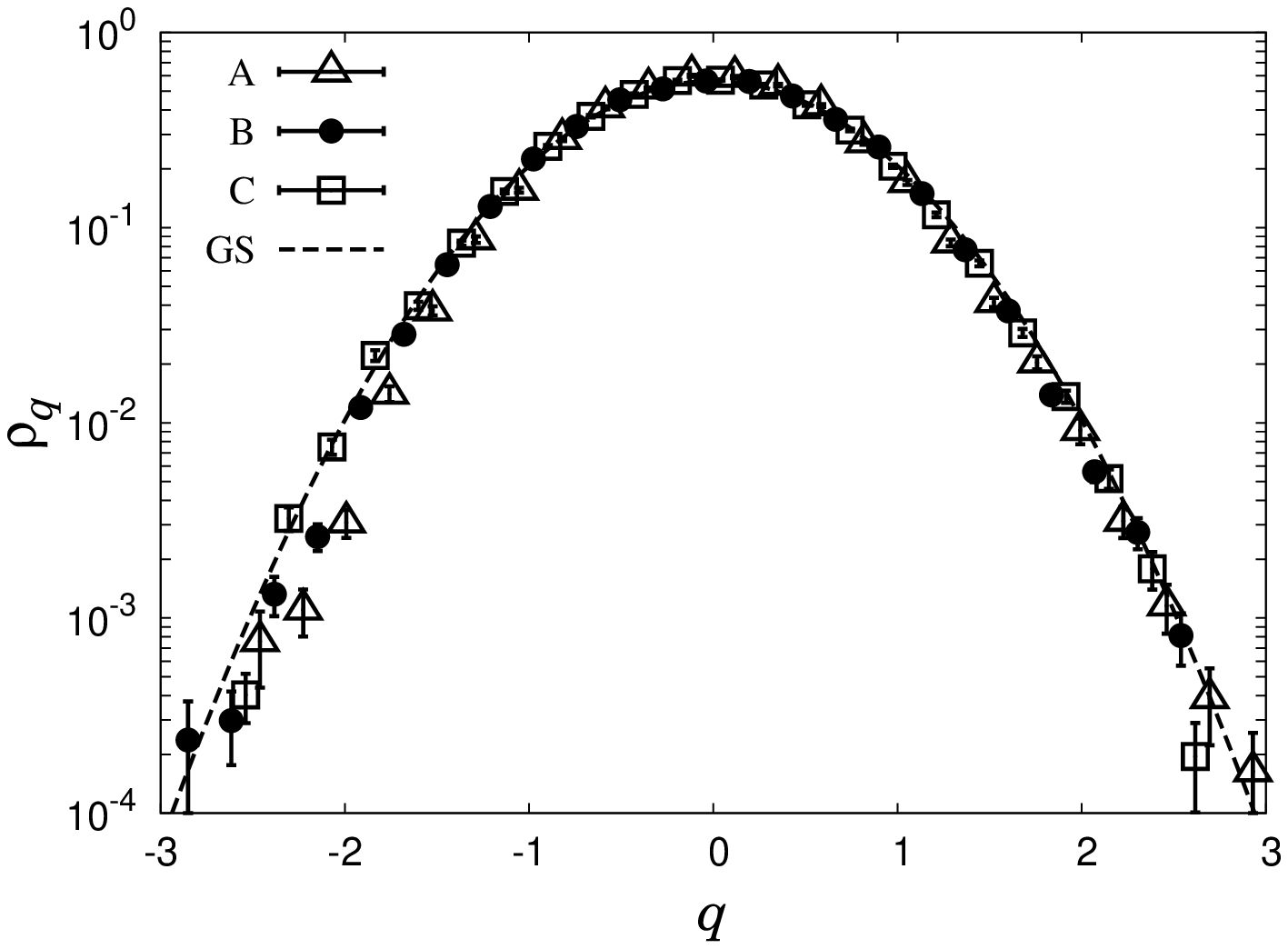} 
  \end{tabular}
  \caption{(Upper) Examples of quantum fluctuations for the harmonic
    oscillator in the conditions A, B, and C (see text).  (Lower)
    Distributions of the coordinate for the harmonic oscillator in
    logarithmic scale together with the squared wave function of the
    ground state denoted by dashed line. }
  \label{fig:HO_WF}
\end{figure}

In order to see the quantum fluctuations in detail, we show in
Fig.~\ref{fig:HO_WF}(upper) examples of the quantum fluctuations out
of the $N=400$ paths.  From the figure, in all conditions the paths
fluctuate from the origin $q=0$ to their maximal amplitude $\sim \pm
1.5$.  The mean squared radius of the fluctuation is
$\Expect{\overline{q^{2}}} = 0.426 \pm
0.08$ 
for the condition A, $0.474 \pm 0.008$ 
for B, and $0.482 \pm 0.008$ for C, 
which are close to the ground-state value $\Expect{q^{2}}_{\rm
  GS}=0.5$ within smaller than about $10 \%$ discrepancies.  We note
that in all conditions fluctuations in large scale are mainly composed
of peak structures of width $\gtrsim 1$.  Especially in the conditions
B and C the fluctuation structures in large scale are described by sum
of several Gauss functions, although they have small fluctuations in
fine scale. Then let us visualize degree of the quantum fluctuations.
For this purpose we make a histogram for $q$ with division of time
range into sufficiently many parts in each path and then combine
$N=400$ histograms to obtain the $q$-distribution $\rho _{q}$.  The
result is shown in Fig.~\ref{fig:HO_WF}(lower) together with the
squared wave function of the ground state.  As one can see, our
$q$-distributions in all conditions behave consistently with the
squared wave function.  Especially it is interesting that behavior of
the quantum fluctuations to large $q$ ($\sim \pm 3$) is very similar
to the squared wave function.

In the examination of our approach for the harmonic oscillator, we
have seen that our approach reproduces quantum behavior of the system
with more than about $90 \%$ accuracy by properly chosen scale
constants, both in random and fixed scale cases.  Especially, by using
smaller scaling constant $\xi$, quantum behavior with higher accuracy
is obtained.


Next let us apply our approach to relativistic field theories.  It is
important that our approach has possibilities to become a
nonperturbative way to quantum field theories. To be specific, we
firstly consider $\UN{1}$ gauge field $A_{\mu}(x)$ in four dimensions
[$x=(\bm{x}, \, \tau )$, $\mu=1$, $2$, $3$, $4$], and assume a
periodic boundary condition with box size (${\cal X}$, ${\cal X}$,
${\cal X}$, ${\cal T}$).  In a similar manner to the nonrelativistic
particles, the quantum transition amplitude of the field in Euclidean
time can be expressed as,
\begin{equation}
{\cal Z} = \int _{\rm period} {\cal D} A \exp ( - S [ A ] ) ,  
\quad 
{\cal D} A \equiv \prod _{x, \mu} 
d A_{\mu} ( x ) .
\end{equation}
In the field path integrals, sole difference to the particle case is
that the field is a function of four components of the coordinate $x$
rather than time $\tau$ only.  Therefore, smooth quantum fields can be
obtained by applying our approach (steps 1--4.) with an extension of
$\tau \to x$.  Then the field fluctuations are expressed as,
\begin{equation}
  A_{\mu} ( x ) = \sum _{i_{\mu}}^{N_{\rm sum}}
  A_{i_{\mu}} \exp \left [ - 
    \frac{(x - x _{i_{\mu}})^{2}}{\xi _{i_{\mu}} ^{2}} \right ] , 
\end{equation}
for each $\mu$, where $(x - x_{i_{\mu}})^{2}$ means to take squared
distance between $x$ and $x_{i_{\mu}}$ in the periodic boundary
condition.

Now let us evaluate quantum fluctuations of the $\UN{1}$ and
$\SUN{2}$ gauge fields in four dimensions, which actions are, 
\begin{equation}
S = \int d^{4}x {\cal L} ( x ), 
\end{equation}
with Lagrangian densities, 
\begin{align}
& {\cal L}_{\UN{1}} = 
\frac{1}{4} \sum_{\mu , \nu =1}^{4} 
( \prt _{\mu} A_{\nu} - \prt _{\nu} A_{\mu} ) ^{2} , 
\\
& {\cal L} _{\SUN{2}} 
= \frac{1}{4} 
\sum _{a=1}^{3} 
\sum_{\mu , \nu =1}^{4}
\Bigg ( 
\prt _{\mu} A^{a}_{\nu} - \prt _{\nu} A^{a}_{\mu} 
+ g \sum _{b,c} \epsilon _{abc} A^{b}_{\mu} A^{c}_{\nu} 
\Bigg ) 
^{2} , 
\end{align}
respectively.  In this study we do not include gauge fixing terms nor
the Faddeev-Popov ghosts in the Lagrangian densities.  The $\SUN{2}$
gauge field has self-interactions with coupling $g$, for which we take
$g=3.5$.  In both $\UN{1}$ and $\SUN{2}$ gauge theories, we take the
following simulation condition.  Namely, we fix the coordinate
$x_{i_{\mu (a)}}$ to be on sites of four-dimensional lattice
$7^{3}\times 14$ dividing the box in same intervals with ${\cal T}=2
{\cal X}$.  We take the fixed scale with $\xi = {\cal X}/(7
\sqrt{\pi}) = {\cal T}/(14 \sqrt{\pi})$.  We note that $\xi$
corresponds to the minimal scale of quantum field theories, as the
lattice spacing $a$ in the discretized framework.  The cut-off for the
fluctuation amplitude, $\Lambda _{A}$, is fixed as $\Lambda _{A} =
1.3~\xi^{-1}$.  At first of the iteration $A_{i_{\mu (a)}}$ is
randomly generated as a hot start. We prepare $N=50$ paths for the
$\UN{1}$ and $\SUN{2}$ gauge fields, respectively.  During cooling by
the Metropolis method, the action of the $\UN{1}$ [$\SUN{2}$] gauge
field converges at around $N_{\rm iteration}\approx 3 \times 10^{5}$
($10^{6}$).  It is interesting that at the saturation point
$\Expect{\overline{\cal L}_{\SUN{2}}} \approx 0.49 ~\xi^{-4}$ is
smaller than $3 \times \Expect{\overline{\cal L}_{\UN{1}}} \approx 3
\times 0.20 ~\xi^{-4}$ due to the self-interactions in $\SUN{2}$,
where $\overline{\cal L}$ is averaged Lagrangian density
[$\overline{\cal L}\equiv \int d^{4}x {\cal L}(x)/{\cal TX}^{3}$].

\begin{figure}[t]
  \centering
  \begin{tabular}{c}
    \Psfig{8.6cm}{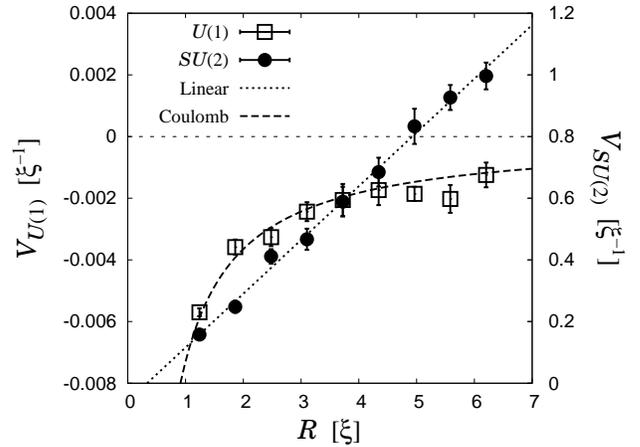} 
  \end{tabular}
  \caption{Potential between fundamental representations for $\UN{1}$
    (left axis) and $\SUN{2}$ (right axis) gauge fields. The $\UN{1}$
    potential is negatively shifted so as to fit the Coulomb
    potential $V(R) = - \alpha / R$ with $\alpha \approx 1/137$
    denoted by dashed line.  Dotted line denotes the linear potential
    $V(R) = \sigma R + b$ fitted to the $\SUN{2}$ potential.  }
  \label{fig:Gauge_potential}
\end{figure}

Quantum fluctuations of gauge fields provide a potential between
(infinitely heavy) fundamental representation and its antiparticle,
which can be evaluated through the Wilson loop of rectangle $C=T
\times R$ defined as~\cite{Wilson:1974sk},
\begin{equation}
W ( T, \, R ) = \trace {\cal P} \exp \left [ 
i g \oint _{C} \sum _{\mu , a} d x_{\mu} A_{\mu}^{a} (x) T^{a} 
\right ] , 
\end{equation}
where ${\cal P}$ means to take the ordered exponential with the group
generator $T^{a}$.  We choose the gauge coupling for $\UN{1}$ as $g =
0.303$ so that $\alpha \equiv g^{2}/(4\pi) \approx 1/137$.  From the
expectation values of the Wilson loop $\Expect{W}$, the potential is
evaluated as,
\begin{equation}
V ( R ) = \frac{1}{t} 
\ln \frac{\Expect{W (T, \, R)}}{\Expect{W (T + t, \, R)}} , 
\end{equation}
for sufficiently large $T$ and small $t$.  In this study, in order to
have enough statistics, we calculate average of $10$ Wilson loops at
random positions in four-dimensional time-space for each set ($T$,
$R$) in each path and then combine the results of $N=50$ paths.  The
results for $\UN{1}$ and $\SUN{2}$ gauge fields are shown in
Fig.~\ref{fig:Gauge_potential}.  As one can see, while the $\UN{1}$
gauge field qualitatively reproduces the Coulomb force, the potential
from the $\SUN{2}$ gauge field shows confining linearity.  A confining
potential is the expected nonperturbative property of non-Abelian
gauge theories inspired by the absence of free quarks in
experiments~\cite{Nakamura:2010zzi}, and our approach implies that
quantization indeed generates confining field configurations in the
$\SUN{2}$ gauge theory.  Fitting our $\SUN{2}$ potential with $\sigma
R + b$ we obtain the string tension $\sigma = 0.17 \pm 0.01 ~\xi
^{-2}$, 
which brings a scale to the quantum $\SUN{2}$ gauge theory.

\begin{figure}[t]
  \centering
  \begin{tabular}{c}
    \Psfig{8.6cm}{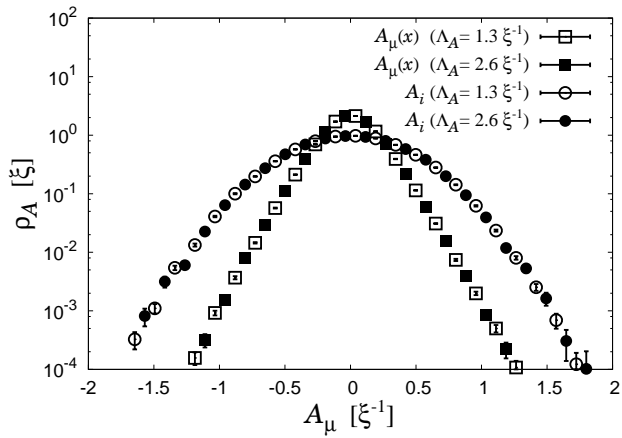} 
  \end{tabular}
  \caption{ Distribution of the $\UN{1}$ field value $A_{\mu}
      (x)$ (open and closed circles) and distribution of the
      coefficient of the Gauss function for the $\UN{1}$ field (open
      and closed squares), with different cut-offs for the fluctuation
      amplitude, $\Lambda _{A}$, in logarithmic scale.  }
\label{fig:Adist}
\end{figure}

Here we should discuss the gauge symmetry in our approach.  In the
path integral formulations of gauge theories without gauge fixing
terms nor Faddeev-Popov ghosts, the measure ${\cal D} A$ and the
Lagrangian ${\cal L}$ are respectively gauge invariant.  This
indicates that numerical simulations in such a condition will have the
gauge symmetry if one takes into account all the possible paths
required by the measure, that is, the full gauge group manifold, which
includes all of the gauge copies and especially all of the Gribov
regions in $\SUN{2}$~\cite{Gribov:1977wm}.  In our approach, however,
we take into account not all but most of the possible paths, as
discussed above, and hence our approach does not contain the full
gauge group manifold.  Then we have two factors which leads to gauge
symmetry breaking; one is that number of the Gauss functions $N_{\rm
  sum}$ is finite and hence lack of the some paths may take place, and
the other is the cut-off for fields, $\Lambda _{A}$, because we
neglect regions out of the cut-off in the simulations.  The first
factor will become unimportant and negligible when one uses large
value of $N_{\rm sum}$ so that distribution of the Gauss functions is
sufficiently dense.  As for the second factor, on the other hand,
regions out of the cut-off might contribute to the path integrals.  In
order to see the cut-off dependence for our results, we simulate
quantum fluctuations of the $\UN{1}$ and $\SUN{2}$ gauge fields with
the cut-off $\Lambda _{A}=2.6 \xi ^{-1}$, which is twice larger than
that in our preceding simulation.  We check that the cut-off
dependence of the expectation values of the Lagrangian densities and
potentials between fundamental representations for both $\UN{1}$ and
$\SUN{2}$ fields is negligible.  Also we plot in Fig.~\ref{fig:Adist}
the distributions of the $\UN{1}$ field value $A_{\mu} (x)$ and of
$A_{i_{\mu}}$, the coefficient of the Gauss function, with two
cut-offs $\Lambda _{A}=1.3 \xi ^{-1}$ and $\Lambda _{A}=2.6 \xi
^{-1}$.  Here the distributions are evaluated in similar manners to
the distribution of the coordinate for the harmonic oscillator
discussed above.  From the figure, both $A_{\mu} (x)$ and
$A_{i_{\mu}}$ dominantly distribute around zero while the
distributions for large values are suppressed at least exponentially
because of the weight $\exp (-S)$ in the simulations.  Due to the
exponential suppression, contributions of the gauge field
configurations to the path integrals are effectively taken into
account if one uses sufficiently large cut-off, which is the case in
our simulations, and as a consequence the cut-off dependence cannot be
seen for both $A_{\mu} (x)$ and $A_{i_{\mu}}$ in Fig.~\ref{fig:Adist}.
This is the result for the $\UN{1}$ gauge field, but we also obtain a
similar result for the $\SUN{2}$ gauge field.  In this sense, we
consider an effective gauge group manifold than the full gauge group
manifold, and in principle the effective manifold can be taken as
close to the full manifold as possible by considering dense
distribution of Gauss functions with large cut-off $\Lambda _{A}$.
Furthermore, this fact leads to a conjecture that in general quantum
fluctuations of gauge fields appear dominantly within certain band and
fluctuations out of the band is suppressed due to the weight $\exp
(-S)$.

In summary, we have developed an approach to evaluation of the
continuum path integrals, in which paths are described by sum of
smooth functions with weight $\exp (-S)$ by the Metropolis method.  In
this study we take an approximation that smooth fluctuations are
described only by the Gauss function.  The weighted smooth paths
reproduce properties of the ground-state harmonic oscillator in one
dimension with more than about $90 \%$ accuracy by properly chosen
width of the Gauss functions.  We have found that quantum behavior
with higher accuracy is obtained by using smaller width of the Gauss
functions, with which finer structure of the quantum fluctuations can
be described.  We have also evaluated quantum fluctuations of fields
and the Coulomb force and confining linear potential have been
extracted at qualitative levels from the $\UN{1}$ and $\SUN{2}$ gauge
fields in four dimensions, respectively.  We have found that
distributions of large values of the gauge fields are suppressed at
least exponentially in our approach, which implies that contributions
of the gauge field configurations to the path integral are effectively
taken into account by sufficiently large cut-off for the fluctuation
amplitude.

This work is partly supported by the Grand-in-Aid for Scientific
Research~(No.~22-3389).  The author acknowledges support by the
Grant-in-Aid for JSPS Fellows.

\end{document}